\documentstyle[11pt]{article}
\begin{document}


\newcommand{\Bbb}{\textrm{Bbb}}

\begin{flushright}
\renewcommand{\textfraction}{0}
July 21th 1994\\
hep-th/9407136\\
PEG-03-94\\
\end{flushright}

\begin{center}
{\LARGE {\bf Strings and Loops in Event Symmetric Space-Time } }
\end{center}
\begin{center}
{\Large Phil Gibbs} \\
\end{center}

\begin{abstract}
Open and Closed super-string field theories are constructed in an
event-symmetric target space. The partition functions of Statistical and
Quantum models are constructed in terms of invariants defined on Lie-algebra
representations. An attractive feature of the closed string models is the
elegant unification of the space-time symmetries with the gauge symmetries.
\end{abstract}

{\it ``It's clear that there is a very deep and beautiful mathematical
structure that underlies all the startling results that we are finding and that
some very elegant and profound principle is there to be found.
... A concern one might have is that the mathematics which is required just
gets so difficult that the human mind is unable to deal with it!''}\\
            John Schwarz

{\it ``In string theory there aren't four or ten dimensions. That's only an
approximation. In the deeper formulation of the theory the whole notion
of what we mean by a dimension of space-time will have to be altered.
... once the correct fundamental formulation of the theory is really
understood, it will probably be something startlingly simple.''}\\
            Michael Green.

Quotes from ``Superstrings: A Theory of Everything?'' \cite{DaBr88}.


\section*{String Theories}

Despite the lack of experimental data above the Electro-Weak energy scale, the
search for unified theories of particle physics beyond the standard model has
yielded many mathematical results based purely on constraints of high symmetry,
renormalisability and cancellation of anomalies. In particular, space-time
supersymmetry \cite{WeZu74} has been found to improve perturbative
behaviour and to bring the gravitational force into particle physics. One
ambitious but popular line of research is superstring theory
\cite{GrSc81a,GrSc81b,GrSc82}. String models were originally constructed
in perturbative form and were found to be finite at each order but incomplete
in the sense that the perturbative series were not Borel summable
\cite{GrPe88}.

There has been some notable success in formulating both open \cite{Wi86} and
closed \cite{Zw92} String Field Theories.
There have also been some important steps taken towards background independent
formulations of these theories \cite{Wi92,SeZw93}.
However, these formulations fail to provide an elegant unification of
space time diffeomorphism symmetry with the gauge symmetry. This is a
significant failure because string theories are supposed to unify gravity
with the other gauge forces.

A successful theory of Quantum Gravity should describe physics at the
Planck scale \cite{Pl99}. It is likely that there is a phase
transition in string theories at their Hagedorn temperature near $kT =$ Planck
Energy \cite{Ha68}. It has been speculated that above
this temperature there are fewer degrees of freedom and a restoration of a much
larger symmetry \cite{AtWi88,GrMe87,GrMe88,Gr88}.

There are also indications that string theory may be discrete in some
sense at short distance scales. It is possible to calculate exact string
amplitudes from a lattice theory with a non-zero spacing
\cite{KlSu88}. Furthermore there are signs of a scaling
duality which also limits measurement of distances smaller than the Planck
length \cite{KiYa84}.

There is another non-perturbative approach to string theories which
gives important insights. Random Matrix Models in the large $N$ double
scaling limit can be shown to be equivalent to two dimensional gravity, or
equally, string theory in zero dimensional target space. The models can be
extended to a one dimensional target space but not to critical dimension where
string theory is perturbatively finite.

One interpretation of the present state of string theories is that it
lacks a geometric foundation and that this is an obstacle to finding
its most natural formulation. It is possible that our concept of
space-time will have to be generalised to some form of ``stringy space''.
Perhaps such space-time must be dynamical and capable of undergoing
topological or even dimensional changes
\cite{AsGrMo93,AnFeKo94}.


\section*{The Loop Representation of Quantum Gravity}

Recently there has been some progress in attempts to quantise Einstein
Gravity \cite{Ei15a,Ei15b} by canonical methods. A reformulation of
the classical theory in which the connection takes the primary dynamic
role instead of the metric \cite{As86,As87} has led
to the Loop Representation of Quantum Gravity \cite{RoSm88,RoSm90}.
The fact that Einstein Gravity is non-renormalisable is considered to be not
necessarily disastrous since gravity theories in 1+1 and 2+1 dimensions have
been successfully quantised by various means \cite{Ca93}.

A similarity between the loop representation and string theories is that
there are attempts to understand them in terms of groups defined on loop
objects \cite{GaTr81}.
This and other similarities between the Loop Representation of Quantum Gravity
and formulations of String Field Theories may be more than superficial
\cite{Ba93}. Superstring theory and the Loop Representation can not
be equivalent since the former only works in ten or eleven dimensions while
the latter only works in four. It is possible that they could be different
phases of the same pre-theory provided that pre-theory allows changes of
dimension.

One other notable aspect of the Loop Representation is that it has a
discrete
nature at scales smaller than the Planck length. The loop area is quantised
in multiples of the Planck length squared. This has inspired renewed
interest in discrete methods.


\section*{Event Symmetric Space-Time}

Whichever approach to quantum gravity is taken the conclusion seems to be that
the Planck length is a minimum size beyond which the Heisenberg Uncertainty
Principle \cite{He38} prevents measurement \cite{Ga94}.
Space-time may have
to be viewed very differently to understand physics beyond the Planck scale.
Until the geometric principles have been understood a consistent
formulation of quantum gravity may be impossible. The increase of
mathematical sophistication in physics which has emerged from these lines
of research may suggest to some that the required mathematics to solve the
problem has not yet been developed. However, the view taken here is that
a discrete approach using a simple geometric principle may provide the
solution.

In the light of what quantum gravity seems to have to say about the
structure of space-time on small scales Wheeler and others have
speculated that a pre-geometry theory is needed \cite{Wh84}. In such a
theory the space-time continuum would not be part of the fundamental
formulation but would arise as a consequence of dynamics. Pre-geometry may
have to have a discrete formulation.

The paradism of Event Symmetric space-time is one such discrete approach
\cite{Gi94a}.  The exact nature of space-time
in this scheme will only become apparent in the solution. Even the number of
space-time dimensions is not set by the formulation and must by a dynamic
result. It is possible that space-time will preserve a discrete
nature at very small length scales. Quantum mechanics is reduced to a
minimal form. The objective is to find a statistical or quantum definition of
a partition function which reproduces a unified formulation of known and
hypothesised
symmetries in physics and then worry about states, observables and causality
later.

We can seek to formulate a lattice theory in which diffeomorphism invariance
takes a simple and explicit discrete form. At first glance it would seem that
only translational invariance can be adequately represented in a discrete form
on a regular lattice but this overlooks the most natural generalisation of
diffeomorphism invariance in a discrete system. Diffeomorphism invariance
requires that the action should be symmetric under any differentiable
1-1 mapping on a $D$ dimensional manifold. This is
represented by the diffeomorphism group $diff(M_D)$. On a discrete
space we  could demand that the action is symmetric under any permutation
of the discrete space-time events ignoring continuity altogether. Generally we
will use the term {\it Event Symmetric} whenever an action has an invariance
under the Symmetric Group $S(\infty)$ over an infinite number of discrete
``events'' (or some larger group of which it is a sub-group).

More precisely the symmetric group is the group of all possible 1-1 mappings
on a set of events. The cardinality of events on a manifold of any number
of dimensions is $\aleph_1$. The number of dimensions and the topology of the
manifold is lost in an event symmetric model since the symmetric groups for
two sets of equal cardinality are isomorphic.

Event symmetry is larger than the diffeomorphism invariance of continuum
space-time.
\begin{equation}
                 diff(M_D) \subset S(M_D) \sim S(\aleph_1)
\end{equation}
If a continuum is to be restored there must be a mechanism of spontaneous
symmetry breaking in which event symmetry is replaced by a residual
diffeomorphism invariance. The mechanism will determine the number of
dimensions of space. It is possible that a model could have several phases
with different numbers of dimensions and may also have an unbroken
event-symmetric phase.

It is unlikely that there would be any way to distinguish a space-time with an
uncountable number of events from space-time with a dense covering of a
countable number of events so it is acceptable to consider models in which the
events can be labelled with positive integers. The symmetry group $S(\aleph_1)$
is replaced with $S(\aleph_0)$. In practice it may be necessary to regularise
to a finite number of events $N$ with an $S(N)$ symmetry and take the large $N$
limit while scaling parameters of the model as functions of $N$.

Renormalisation and the continuum limit must also be considered but it is not
clear what is necessary or desired as renormalisation behaviour. In quantum
field theories with a
lattice formulation such as QCD it is normally assumed that a continuum limit
exists where the lattice spacing tends to zero as the renormalisation group
is applied. In string theories, however, the theory is perturbatively finite
and the continuum limit of a discrete model cannot be reached with the
aid of renormalisation. It is possible that it is not necessary to have an
infinite density of events in space-time to have a continuum or there may
be some alternative way to reach it, via a q-deformed non-commutative geometry
for example \cite{DiMu94a}.

It stretches the imagination to believe that a simple event symmetric model
could be responsible for the creation of continuum space-time and the
complexity of quantum gravity through symmetry breaking, however, nature has
provided some examples of similar mechanisms which may help us accept the
plausibility of this claim.

The simplest possibility is to model space-time as a critical solid
formed from randomly bonded points \cite{Or93}. The points are assigned
a set of $D$ real numbers and are analogous to molecules moving in a $D$
dimensional space. For a suitable action symmetric in exchange of molecules
they can model a critical solid at a
second order melting phase transition. This gives rise dynamically to what
might be interpreted as a $D$ dimensional curved manifold. In this case
the number of dimensions is predetermined and it is difficult to see how the
space-time could form different topologies.

There is another variant of this natural mechanism that has more flexibility.
Consider the way in which soap bubbles arise from a statistical physics
model of molecular forces. The forces are functions of the relative
positions and orientations of the soap and water molecules. The energy is
a function symmetric in the exchange of any two molecules of the same
kind. The system is consistent with the definition of event symmetry since it
is invariant under exchange of any two water or soap molecules and therefore
has an $S(N) \otimes S(M)$ symmetry where $N$ and $M$ are the number of water
and soap molecules. Under the right conditions the symmetry breaks
spontaneously to leave a diffeomorphism invariance on a two dimensional
manifold in which area of the bubble surface is minimised.

Events in these
models correspond to molecules rather than space-time points. Nevertheless,
they
are perfect mathematical analogies of event-symmetric systems where the
symmetry breaks in the Euclidean sector to leave diffeomorphism invariance in
two dimensions as a residual symmetry. Indeed the models illustrate a deep
analogy between events in event symmetric space-time and many-particle systems.
The models considered further are more sophisticated than the molecular
models and do not predetermine geometry in any way. However, the analogy
between particles and space-time events remains useful.

A number of Event Symmetric models will be described in this paper. Some of
these can best be understood as statistical theories with a partition function
defined for a real positive definite action.
\begin{equation}
                  {\Large Z = \int e^{-S} }
\end{equation}
Others can only be considered as quantum theories for which the action need
not be positive definite provided the partition function is well defined
\begin{equation}
                  {\Large Z = \int e^{iS} }
\end{equation}
It is not always clear when such an integral should be considered well
defined. For example the action,
\begin{equation}
                   S = x^2 - y^2
\end{equation}
gives a well defined quantum partition function in the two variables $(x,y)$
but if the variables are transformed by a 45 degree rotation to $(u,v)$, the
action becomes
\begin{equation}
                   S = 2uv
\end{equation}
for which the integral is not well defined.

It might be safer to consider only positive definite actions and assume that
in a physically valid theory, the only difference between the statistical
event symmetric model and the quantum one should be a factor of $i$ against
the action in the exponential. We might expect that in the statistical version
the Event Symmetry will break to give Riemannian space-time with a Euclidean
signature metric while in the quantum version it breaks to give the physical
Lorentzian theory with Minkowski signature metric.

It is not clear if this is realistic, after all, continuum lagrangian densities
for field theories in Lorentzian space-time are made non-positive definite by
the signature of the metric. It is not clear what conditions should be placed
on the form of an event-symmetric action to ensure a well defined tachyon free
quantum theory which produces dynamically the correct Lorentz signature. Even
in continuum theories this is an interesting question and it is believed that
a Lorentzian signature is preferred for certain theories in 4 and 6 dimensions.
\cite{CaGr93}.


\section*{Random Matrix Models}

A basic type of event-symmetric model places field variables $A_{ab}$ on links
joining all pairs of events $(a,b)$. A suitable action must be a real scalar
function of these variables which is invariant under exchange of any two
events.

The link variables $A_{ab}$  can be organised into the upper
triangle of a matrix. If there are no self links the diagonal terms are
zero so it is natural to extend the matrix to the lower half by making the
matrix anti-symmetric.

A four link loop action is
\begin{equation}
               S = \sum_{a,b,c,d} A_{ab} A_{bc} A_{cd} A_{da}
\end{equation}

This is equal to
\begin{equation}
               S = Tr(A^4)
\end{equation}

which is an invariant under $O(N)$ similarity transformations on the matrix.

This suggests that we consider actions which are functions of the traces
of powers of the matrix $A$. Then the symmetry group of the system is $O(N)$
which has $S(N)$ as a subgroup. The idea can be extended to unitary groups
by using complex variables for hermitian matrices or symplectic groups
by using quarternions.

This is an appealing idea since it naturally unifies the $S(N)$ symmetry,
which we regard as an extension of diffeomorphism invariance, with gauge
symmetries. If the symmetry broke in some miraculous fashion then it
is conceivable that the residual symmetry could describe quantised gauge
fields on a quantised geometry. For example a discrete gauge $SO(10)$ symmetry
on a lattice of M points would be,
\begin{equation}
               SO(10)^N \subset O(10 N)
\end{equation}

For the symmetry to break in the way we desire, i.e.
leaving a finite dimensional topology, the events
will have to organise themselves into some arrangement where there is an
approximate concept of distance between them perhaps defined by correlations
between field variables. Matrix elements linking events which are
separated by large distances would have to be correspondingly small. Only
variables which are localised with respect to the distance could have
significant values.

There are several
possible generalisations to multi-matrix models, tensor models and models
with fermions. In each case the action can be a function of any set of
scalars derived from the tensors by contraction.
For the action to reduce to an effective local action on this space the
original action must be restricted to forms in which it is the sum of
terms which are written as contractions over tensors and which do not
separate into products of two or more such scalar quantities. For example
if there are two matrices A and B defining the field variables then
the action could contain terms such as,
\begin{equation}
                        tr(ABAB)
\end{equation}
but not,
\begin{equation}
                       tr(AB)^2
\end{equation}
or
\begin{equation}
                       tr(A)tr(B)
\end{equation}
This locality condition is important when selecting suitable actions for
models which might exhibit dimensional symmetry breaking.

This type of random matrix model has been extensively studied in the
context where $N$ is interpreted as the number of colours or flavours. (see
\cite{ItDr89,Kak91} ) The event-symmetric paradism suggests an alternative
interpretation in which $N$ is the number of events times the number of
colours.

Symmetry breaking in one-matrix models appears to be limited to simple forms
\cite{BrDeJaTa92}. One interesting result
is that the perturbation theory of an $SU(N)$ matrix model in the large $N$
double scaling limit is equivalent to a $c=0$
string theory \cite{tH74,Ka89}.

For the action to reduce to an effective local action on this space the
original action must be restricted to forms in which it is the sum of
terms which are written as contractions over tensors and which do not
separate into products of two or more such scalar quantities. For example
if there are two matrices A and B defining the field variables then
the action could contain terms such as,
\begin{equation}
                       tr(ABAB)
\end{equation}
but not,
\begin{equation}
                       tr(AB)^2
\end{equation}
or
\begin{equation}
                       tr(A)tr(B)
\end{equation}

This locality condition is important when selecting suitable actions for
models which might exhibit dimensional symmetry breaking.

However, it seems likely that this type of model cannot break event symmetry in
any useful way given the invariance and locality conditions described if there
are a finite number of tensors involved.


\section*{Event Supersymmetric Matrix Models}

It would be an obvious next step to generalise to supersymmetric matrix models
\cite{MaPa90}. So far we have matrix models based a families of
groups such as $S(N)$, $O(N)$,
$SU(N)$ or $Sp(N)$. Tensor representations and invariants can be used to
construct models with commuting variables, anticommuting variables or
both. Similarly we can define models based on supersymmetry groups of
which there are also several families such as $SU(N/M)$ and $OSp(N/M)$.
For analysis of supergroups see \cite{Co89}

Just one simple model will
be described. The representation has an anti-hermitian matrix
$A$ of commuting variables
\begin{equation}
                   A^*_{ab} = -A_{ba}
\end{equation}
and a vector $\psi$ of anti-commuting variables.
A suitable action could be,
\begin{equation}
 S = m(2 i \psi^*_a \psi_a + A_{ab} A_{ba})\\
     + \beta( 3 \psi^*_a A_{ab} \psi_b -
        i A_{ab}A_{bc}A_{ca} )
\end{equation}
As well as $U(N)$ invariance this is invariant under a super-symmetry
transform with an infinitesimal anticommuting parameter $\epsilon_b$,
\begin{equation}
 \delta A_{ab} = \epsilon^*_b \psi_a - \epsilon_a \psi^*_b\\
	  \delta \psi_a = i \epsilon_b A_{ab}\\
	  \delta \psi^*_a = i \epsilon^*_b A_{ba}\\
\end{equation}

It is encouraging that supersymmetric generalisations of matrix models
can be so easily
constructed on event symmetric space-time. Demanding supersymmetry helps
reduce our choice of actions
but not actually very much. There are still many different possibilities
like the above which can be constructed from contractions over tensor
representations of supersymmetry groups. These models are special cases of
matrix or tensor models so they will not be more successful
as a scheme for dimensional symmetry breaking.


\section*{Event Symmetric String Models}

The fact that a large number of degrees of freedom are perhaps required to
produce event symmetry breaking suggests that string theories might
provide answers. The place to start is with
string groups \cite{Ka88}.

To define the string groups an open orientated string is first considered as
a topological object independently of any target space. There is an abstract
operator $L_C$ for each string $C$ and a Lie-product is defined for these
operators by specifying the structure constants,
\begin{equation}
                 L_A \wedge L_B = \sum f^C_{AB} L_C
\end{equation}
Three strings $(A,B,C)$ are said to form a triplet if the end of A matches
the start of B, the end of B matches the start of C and the end of C matches
the start of A. They must match in such a way that there is no part of any
string unmatched. In other words they form three matching lengths
radiating from one central point.

When $(A,B,C)$ is such a triplet then the structure constants are
\begin{equation}
                 f^{C^T}_{AB} = f^{A^T}_{BC} = f^{B^T}_{CA} = 1
\end{equation}
and
\begin{equation}
                 f^{C^T}_{BA} = f^{A^T}_{CB} = f^{B^T}_{AC} = -1
\end{equation}
$C^T$ is the transposed string formed by reversing its orientation.

All other structure constants are zero. It can be checked that this
does define a Lie-algebra because the product is anti-symmetric
and satisfies the Jacobi identity. This Lie-algebra is regarded as generating
the gauge group of the open string field theories.

The algebra can be realised on a continuous $D$ dimensional manifold $M_D$
when the strings are continuous open orientated curve segments in the space.
The group is then called the Universal String Group $str(M_D)$

Such string groups have been used to formulate both open and closed string
field theories \cite{Kak91}.

These formulations are, however, not completely satisfying. If string
theory really unifies gravity with the gauge forces then the symmetry
group of gravity, $diff(X)$ on a manifold $X$, should be unified with the
string group, $str(X)$ on target space X. This is not achieved in the
continuum string field
theories. Furthermore the phenomenon of topology change suggests that it
is not possible \cite{Wi93} because the string groups must be
the same on topologically different manifolds.
Ideally the diffeomorphism groups should be contained in the full string
groups
\begin{equation}
                 diff(X) \subset str(X) \sim str(X')
\end{equation}
But then the string group must contain $diff(X)$ for every possible
topologically different manifold X.

This seems quite unreasonable but in fact it is exactly what happens
in event symmetric field theories which contain the full event symmetric
group $S(X)$. These groups are isomorphic for any two manifolds and contain
the diffeomorphism groups
\begin{equation}
                 diff(X) \subset S(X) \sim S(X')
\end{equation}
The solution will be to find a string group which contains the symmetric group
\begin{equation}
                 diff(X) \subset S(X) \subset str(X)
\end{equation}
We shall see how it is possible to define string groups on an event
symmetric target space of discrete points in such a way that the
symmetric group is included as a subgroup.


\section*{Discrete Open String Associative Algebras}

To begin constructing event symmetric string models we will extend
matrix algebras to discrete string algebras then use these to construct
extensions of matrix models \cite{Gi94b}.

A basis for a discrete open string vector space is defined by the set of
open ended oriented strings through an event symmetric space of $N$ events.
E.g. a possible basis element might be written,
\begin{equation}
              C = (1,4,3,1,7)
\end {equation}
Note that a string is allowed to intersect itself. In the example the string
passes through the event $1$ twice. A string must be at least two events
long. A null string passing through zero events or Strings
passing through just one event could be included but are not needed. The
order in which the string passes through the events is significant e.g.
\begin{equation}
                   (1,4,3) \neq (1,3,4)
\end {equation}
but the order in which the events themselves have been numbered is
irrelevant since the models are to be event symmetric.
A complete set of field variables in an event symmetric string model would
be an element of this infinite dimensional vector space which could be written
as a sum over strings $C$
\begin{equation}
                     \Phi = \sum \Phi^C C
\end {equation}
This can define either a real or complex vector space. To avoid questions
about convergence in some of the definitions that follow it is easiest to
specify that only a finite number of the components can be non-zero. Other
ways of regularising could be used or the sums could be regarded as just
formal expressions.

An inner product can be defined.
\begin{equation}
             \Phi_1 \bullet \Phi_2 = \sum \Phi_1^{C*} \Phi_2^C
\end {equation}
[An asterisk is used to denote complex conjugation.]

The inner product will prove useful when we wish to define a positive
definite form for an action since,
\begin{equation}
             \Phi \bullet \Phi = \sum |\Phi^C|^2 \geq 0
\end {equation}

To add the structure of an associative algebra the product $AB$ of two strings
in the space is defined by joining them when the end of $A$ matches the
beginning of $B$ reversed. It is necessary to add together all the ways in
which this can be done e.g.
\begin{equation}
        (1,4,3,1,7)(7,1,5,1) = (1,4,3,5,1) + (1,4,3,1,1,5,1)
\end{equation}
In the case where the last point of the first string is not the first
point of the second the product is always zero. This ensures that
string models are local, i.e. strings which do not intersect should
not interact directly. E.g.
\begin{equation}
                     (1,4,3,1,7)(1,6) = 0
\end{equation}
To ensure associativity this rule must be balanced with another rule that
if the whole of one of the strings in a product matches, the last event
is not cancelled e.g.
\begin{equation}
                     (1,4,3,1,7)(7,1) = (1,4,3,1,1)
\end{equation}
It can now be checked that these rules define an associative multiplication,
and that it is closed over strings of length 2 and greater.
\begin{equation}
                     A(BC) = (AB)C
\end{equation}
The multiplication of strings extends immediately to multiplication on
the vector spaces.
\begin{equation}
		     \Phi_1 (\Phi_2 \Phi_3) = (\Phi_1 \Phi_2) \Phi_3
\end{equation}

The algebra has an identity
\begin{equation}
                     I = \sum_{i=1}^N (i,i)
\end{equation}

Examining the local string algebra it is observed that there is a sub-algebra
spanned by the rank two bases,
\begin{equation}
  (a, b) (c, d) = \delta_{bc} (a, d)
\end{equation}
The sub-algebra is isomorphic to multiplication of $N \times N$ Real or Complex
matrices.

A notation for the string algebras will be adopted which reflects this
relationship between the string algebras and matrix algebras. The string
algebras are extensions of the algebras $M(N,{\Bbb R})$ and $M(N,{\Bbb C})$
and will be written $Open(M(N,{\Bbb R}))$ and $Open(M(N,{\Bbb C}))$.

There is an alternative non-local
algebra in which two strings which do not share common points have a
non-zero product, e.g.
\begin{equation}
                     (1,4,3,1,7)_E (1,6)_E = (1,4,3,1,7,1,6)_E\\
		     (1,4)_E (4)_E = (1)_E + (1,4,4)_E
\end{equation}
In this algebra it is necessary to include a null string and strings of length
one. The null string is the identity. The real and complex algebras on the
vector spaces spanned by these base strings can be
denoted by $Open(N,{\Bbb R})$ and $Open(N,{\Bbb C})$. These non-local (global)
algebras turn out
to be (isomorphic to) sub-algebras of the local ones. This is realised by
summing over strings with first and final events equal.
\begin{equation}
                     (1,4,3,1,7)_E = \sum_{i=1}^N (i,1,4,3,1,7,i)
\end{equation}
On further examination it emerges that the local algebra factorises into
the tensor product of the matrix algebra acting on the two end points
and the global algebra acting on the rest.
\begin{equation}
          Open(M(N,{\Bbb C})) = Open(N,{\Bbb C}) \otimes M(N,{\Bbb C})
\end{equation}
This leads to some more general definitions. Firstly there is no need for
$N$ in either of the two factors to be the same so define,
\begin{equation}
          Open(N,M(L,{\Bbb C})) = Open(N,{\Bbb C}) \otimes M((L,{\Bbb C})
\end{equation}
The open string extension of a general associative algebra {\cal A} can be
defined as,
\begin{equation}
          Open(N,{\cal A}) = Open(N,{\Bbb C}) \otimes {\cal A}
\end{equation}

For the moment we return to the specific case of the extended matrix algebras
to define the generalised trace and adjoints. The trace is best defined as
the trace from the matrix part. I.e. there is a contribution from the
components of length two strings on the matrix diagonal only.
\begin{equation}
                 Tr(C) = 1 if C = (i,i) and = 0 otherwise
\end{equation}
There is an extended trace defined as the sum over components of any
even length string which is palindromic, i.e. the same when reversed e.g.,
\begin{equation}
                   OTr(1,4,4,1) = OTr(3,3) = 1\\
                   OTr(2,3) = OTr(1,2,1) = 0
\end{equation}
Both these traces behave like traces should and in particular,
\begin{equation}
                 Tr(\Phi_1\Phi_2) = Tr(\Phi_2\Phi_1)\\
                 OTr(\Phi_1\Phi_2) = OTr(\Phi_2\Phi_1)
\end{equation}

The orientation reversal of strings will be used to define transposition
denoted with a $T$. e.g.
\begin{equation}
		    (1,5,4)^T = (4,5,1)
\end{equation}
The adjoint of a general element of the space, denoted by a dagger, is defined
by transposing each base
element and in the case of the complex space the complex conjugate of the
components is also taken. I.e.
\begin{equation}
                     \Phi^\dagger = \sum \Phi^{C*} C^T
\end {equation}
The usual relation between adjoints and multiplication holds
\begin{equation}
                     (AB)^T = B^T A^T\\
                     (\Phi_1 \Phi_2)^\dagger = \Phi_2^\dagger \Phi_1^\dagger
\end{equation}
Finally the inner product can be written in terms of these operations.
\begin{equation}
                     \Phi_1 \bullet \Phi_2 = Tr(\Phi_1^\dagger \Phi_2)
\end {equation}


\section*{The Open String Lie-Algebras}

{}From the associative algebra $Open(M(N,{\Bbb C}))$ an infinite dimensional
Lie algebra can be defined with the Lie product being given by the
anticommutator,
\begin{equation}
                     A \wedge B = AB - BA
\end{equation}
This product automatically satisfies the Jacobi identity because of the
associativity of the original algebra product,
\begin{equation}
   A \wedge (B \wedge C) + B \wedge (C \wedge A) + C \wedge (A \wedge B) = 0
\end{equation}
With the wedge product the algebra is an infinite dimensional Lie Algebra
and in principle it defines a group by exponentiation. To avoid complications
in this process only the lie-algebras will be considered.

Using the structure constants for the algebra the Lie product can be written
\begin{equation}
    A \wedge B    =   \sum f^C_{AB} C
\end{equation}
string indices are formally lowered and raised by using the metric to reverse
the direction of the string so that
\begin{equation}
                       f_{CAB} = f^{C^T}_{AB}
\end{equation}
The three strings C, A and B are said to form a triplet is $f_{ABC}$ is
plus one, and an anti-triplet if it is minus one. They are a triplet
if and only if they are all different and the end of A matches the beginning
of B, the end of B matches the beginning
of C and the end of C matches the beginning of A without any events being
left out or used twice. Anti-triplets are triplets with two of the strings
interchanged. It follows that the structure constants are fully antisymmetric.
\begin{equation}
         f_{ABC} = f_{BCA} = f_{CAB} = -f_{ACB} = -f_{CBA} = -f_{BAC}
\end{equation}
The following important relation is also valid
\begin{equation}
        A \bullet ( B \wedge C ) = (A \wedge B) \bullet C = f_{ABC}
\end{equation}
{}From this description of the Lie-product the relationship with the Universal
String Group is clear. The only essential difference is that the group is now
defined on an event-symmetric space-time rather than a continuous one.

Because the Lie-product was defined as the anticommutator on the string
extended matrix algebra $Open(M(N,{\Bbb C}))$ we know that the Lie-algebra
must be an extension of the general linear Lie-algebra. This is confirmed
by the relation,
\begin{equation}
  (a, b) \wedge (c, d) = \delta_{bc} (a, d) -
                                       \delta_{ad} (c, b)
\end{equation}
The algebra is therefore given the name $open(gl(N,{\Bbb C}))$. A number of
other extended Lie-algebras follow immediately by using the anti-commutator
of the appropriate extended associative matrix algebras. e.g.
$open(gl(N,{\Bbb R}))$, $open(gl(N,{\Bbb H}))$, $open(N,gl(L,{\Bbb C})$
and of course $open(N,{\Bbb C})$.

The trace of the matrix algebras can also be used to define the special
subgroups because,
\begin{equation}
                  Tr(A \wedge B) = 0
\end{equation}
The sub-algebra of traceless elements of $open(gl(N,{\Bbb C}))$ will be
denoted by $open(sl(N,{\Bbb C}))$. The Open trace can also be used to
define subgroups. I.e. the elements of $open(gl(N,{\Bbb C}))$ for which
\begin{equation}
                  OTr(\Phi) = 0
\end{equation}
form the sub-algebra $sopen(gl(N,{\Bbb C}))$. There is also an algebra
$sopen(N,{\Bbb C})$ defined in this way and if both traces are used we
have $sopen(sl(N,{\Bbb C}))$.

There is an important alternative definition of the special groups for matrix
algebras. Given a Lie Algebra ${\cal L}_0$ a subalgebra is defined as those
elements which are formed from the Lie-product.
\begin{equation}
                  {\cal L}_1 = {\cal L}_0 \wedge {\cal L}_0
\end{equation}
If ${\cal L}_0$ is $gl(N,{\Bbb C})$ then ${\cal L}_1$ is $sl(N,{\Bbb C})$.
For the string extended algebras there are many linear invariant operators $O$
which have the tracelike property.
\begin{equation}
                  O(\Phi_1 \wedge \Phi_2) = 0
\end{equation}
so applying the same technique to $open(gl(N,{\Bbb C}))$ will give a
sub-algebra of $sopen(sl(N,{\Bbb C}))$. For the present only the traceless
definition will be used.

Of more importance to event symmetric string theories are the open string
extensions to the families of Lie-algebras of compact matrix groups $so(N)$,
$u(N)$ and $sp(N)$. These are easy to define with the adjoint operator which
has the property,
\begin{equation}
     (\Phi_1 \wedge \Phi_2)^\dagger = - (\Phi_1^\dagger \wedge \Phi_2^\dagger)
\end{equation}
The algebra $open( u(N) )$ is defined as the sub-algebra of
$open(gl(N,{\Bbb C}))$ containing all elements for which
\begin{equation}
                  \Phi^\dagger = - \Phi
\end{equation}
The algebras $open( so(N) )$ and $open( sp(N) )$ are the similarly defined
sub-algebras of $open(gl(N,{\Bbb R}))$ and $open(gl(N,{\Bbb H}))$.

There are also compact groups derived from the non-local groups
$Open(N,{\Bbb R})$ etc which will be denoted by $Comp(N,{\Bbb R})$,
$Comp(N,{\Bbb C})$ etc.


\section*{Statistical and Quantum Models for Open Strings}

To define a model or theory which incorporates the group structures defined
in the previous sections we need to choose a representation and an invariant
action. The obvious representation to choose is the fundamental representation
which takes elements of the lie-algebra
\begin{equation}
   \Phi = \sum \phi_{ab} (a, b) + \sum \phi_{abc} (a, b, c) + ...\ldots
\end{equation}
The infinitesimal transformations are generated by an element $\epsilon$ of
the algebra as follows,
\begin{equation}
          \delta \Phi = \Phi \wedge \epsilon
\end{equation}
There are many alternative representations formed from tensor products, direct
sums etc but the fundamental representation has the advantage that there is
exactly one component field variable for each degree of symmetry.

The action should be real and must satisfy a certain locality
principle. It will take a polynomial form in the components of the
representation and in no term must there appear a product of two components
of strings which do not pass through the same event. This rules out the
non-local groups $Open(N,{\Bbb C})$, $Comp(N,{\Bbb C})$ etc since they have
very few invariants which are local in this sense. The special groups will also
be ruled out since constraints such as $Tr(\Phi) = 0$ can be considered
non-local.

The trace is a source of invariants since
\begin{equation}
          \delta Tr(\Phi) = Tr(\Phi \wedge \epsilon) = 0
\end{equation}
Furthermore the associative product can be used since the lie-algebra acts
like a differential operator on the extended matrix algebra according to the
Leibnitz rule,
\begin{equation}
	(\Phi_1 \Phi_2) \wedge \epsilon =
	 (\Phi_1 \wedge \epsilon) \Phi_2 + \Phi_1 (\Phi_2 \wedge \epsilon)
\end{equation}
So there is an infinite sequence of invariants given by,
\begin{equation}
                 {\cal I}_n = Tr(\Phi^n),   (n = 1,\ldots)
\end{equation}
Another sequence of invariants can be defined using the extended trace and
there are many other possible invariants but for simplicity only these
will be considered.
Any action which is written as a sum of these invariants is consistent with
the locality condition.
\begin{equation}
                 S = \sum g_n{\cal I}_n
\end{equation}
A statistical model has a partition function defined on a real action which is
positive definite or at least bounded below. For the string extended general
linear groups the trace invariants are not positive definite. This problem is
resolved in the same way as it is for matrix models by using the Lie-algebras
of the compact groups for which
\begin{equation}
                    \Phi^\dagger = - \Phi
\end{equation}
Then the even trace invariants can be written,
\begin{equation}
          {\cal I}_2n = tr(\Phi^{2 n}) = (-1)^n \Phi^n \bullet \Phi^n
\end{equation}
The simplest non-trivial action for a statistical model is therefore
\begin{equation}
   S  =  m \Phi \bullet \Phi + \beta \Phi^2 \bullet \Phi^2
\end{equation}

[It is important to recognise that the model has  an infinite number of
degrees of freedom even for finite $N$. It would be necessary to demonstrate
that it can give a well defined model despite this.]

There are many other possibilities but this is the most immediately interesting
bosonic open string statistical model. It is also possible to construct
fermionic models using representations such as
\begin{equation}
                 \Psi = \sum \Psi^C C
\end{equation}
Where the components $\Psi^C$ are anticommuting Grassman variables. an action
for this model can be written,
\begin{equation}
   S  =  im \Psi \bullet \Psi  +
   \beta (\Psi \wedge \Psi) \bullet (\Psi \wedge \Psi)
\end{equation}

The extended trace can also be used to define positive definite actions because
$OTr(\Phi^2)$ is bounded even though it contains such non-square terms as,
\begin{equation}
                 \sum \phi_{ab}\phi_{bdda}
\end{equation}

For quantum models the conditions can be relaxed a little since the action does
not have to be positive definite to give a well defined partition function.
The general linear groups are still ruled out but extended Poincare groups
might be considered as well as the compact groups and the odd trace invariants
could also be valid terms in the action.

For open string models there appear to be many possible gauge groups, many
possible representations and many possible invariants. There are several ways
to generate many more possibilities than have been described here. For example
models of charged strings can be constructed from algebras such as
$open( N, so(10 N) )$. Some further criterion would be needed to select a
good theory. It is possible to speculate that only a small number of these
models would have the desired symmetry breaking features to identify them
as good theories. This might be considered unsatisfactory since it would be
better to have a kinematic reason for selecting the right model rather than a
dynamic one.

Another feature of the open string models which is unsatisfactory is that the
event symmetry is not unified with the gauge group. It is true that the
extended matrix models include the symmetric group as a subgroup of the matrix
group. However, true event symmetry is invariance under permutation of events
and although the models above possess this invariance it is not the same as
the symmetric subgroup of the matrix group which acts only on the ends of
the strings.


\section*{Alternative Open String Groups}

before moving on it is necessary to mention some alternative groups which
could be used to construct similar theories to the open string models
described above.

A question that might be asked is ``Is it necessary to use an event
symmetric target space rather than, say, a regular lattice?''. For the
groups as constructed above the answer is that the space must be event
symmetric. If you try to restrict to strings which only follow links
on a regular lattice you find that the group cannot be closed.

However, there is an alternative string group which does close on
any lattice. For this group the associative multiplication rule is
modified to keep the last event common to the two strings. E.g.
\begin{equation}
            (1,2,3,4) (4,3,5) = (1,2,3,4,3,5) + (1,2,3,5)
\end{equation}
This can be used to define models with symmetries very similar to
the models already described. In this case the strings restricted
to follow the links of any lattice form a closed sub-algebra of
the full event-symmetric algebra.

These groups will therefore be called lattice string groups. They have
all the useful properties described for the open string groups but are
not extensions of matrix algebras. These algebras will not be discussed
further.

Another algebra which could replace $Open(N,{\Bbb R})$ has a simpler
multiplication rule in which the strings are simply joined without
adding any terms where part of the string is cancelled. E.g
\begin{equation}
            (1,2,3,4) \circ (5,6,7) = (1,2,3,4,5,6,7)
\end{equation}
This is a non-local algebra but it can be used to define another
class of string extensions for matrix algebras which are local. What makes
it interesting, however, is that the corresponding group on a continuous
target space has recently been identified as interesting in the context of
the loop representation of quantum gravity where it is known as the
Extended Loop Group \cite{BaGaGr93}.

This group is
less suitable for string theories since it does not correspond to the
Universal String Group in the same way as the event symmetric Open string
groups and the lattice groups do.


\section*{Supersymmetric String Groups}

An attractive feature of the discrete string groups on event symmetric
space-time is that supersymmetric versions can be constructed in a very natural
way.

The matrix algebras $M(N, {\Bbb R})$ and $M(N, {\Bbb C})$ can be generalised
to superalgebras $M(L/K, {\Bbb R})$ and $M(L/K, {\Bbb C})$
\cite{Co89}. From these super algebras a number of families of
super Lie-algebras can be constructed of which the most important include
$gl(L/K, {\Bbb R})$, $gl(L/K, {\Bbb C})$, $u(L/K)$, $osp(L/K)$.

It is possible to apply the string extension methods for ordinary algebras
to these superalgebras to construct $Open(N, M(L/K, {\Bbb C}))$,
$open(N, u(L/K))$ etc. This can be improved by first generalising
$Open(N,{\Bbb C})$ to the super-symmetric algebra $Open(L/K, {\Bbb C})$.
To define this algebra it is sufficient to describe a consistent grading
of the base strings into odd and even strings. To do this the events
themselves are given parity so that event-supersymmetric space-time contains
$L$ even events and $K$ odd events. For notational convenience even events will
be labelled with even integers and odd events with odd integers.

The parity of a string is defined to be the total parity of the events it
passes through. The parity
of a string $C$ written $par(C)$ is zero for even strings and one for odd
strings. This defines a grading of the vector space which is
consistent with the associative multiplication since the parity of the
product of two strings is the sum of their parities modulo two
\begin{equation}
                     par(AB) = par(A) + par(B) - 2 par(A) par(B)
\end{equation}
The components of the vectors must be taken from a Grassman algebra
with even (commuting) variables for components of even strings and
odd (anti-commuting) variables for components of odd strings i.e.
\begin{equation}
                      \Phi = \sum \Phi^C C\\
                \Phi^A \Phi^B = (-1)^{par(A)par(B)} \Phi^B \Phi^A
\end{equation}
The real and complex algebras defined in this way are denoted by
$Open(L/K, {\Bbb R})$ and $Open(L/K, {\Bbb C})$. Note that while
$Open(L/0, {\Bbb R})$ is isomorphic to $Open(L, {\Bbb R})$, the
algebra $Open(0/L, {\Bbb R})$ is a super-algebra in which the parity
of a string is the parity of its length. This is in contrast to the
matrix algebras for which $M(L/0, {\Bbb R})$ is the same as
$M(0/L, {\Bbb R})$.

It is now possible to define local super-matrix algebras
$Open(L/K, M(P/Q, {\Bbb C}))$ using the tensor product prescription.
In the case $P=L$ and $Q=K$ we write simply $Open(M(L/K, {\Bbb R}))$ for
consistency the indices of the matrix algebra are also taken as odd and
even.

The adjoint operator must fulfill the usual relation
\begin{equation}
                (\Phi_1 \Phi_2)^\dagger = \Phi_2^\dagger \Phi_1^\dagger
\end{equation}
This is achieved by modifying the previous definition to include a factor of
$i$ when taking the adjoint of an odd element. This restricts us to the
complex version of the model.
\begin{equation}
                     \Phi^{\dagger} = \sum i^{par(C)} \Phi^{C*} C^T
\end{equation}

When generalising the definition of trace and extended trace extra sign factors
are needed corresponding to the parity of half the even string. E.g.
\begin{equation}
                   Tr(2,2) = 1,  Tr(3,3) = -1\\
                   OTr(3,5,5,3) =  1,  OTr(1,4,4,1) = -1
\end{equation}

String extended super Lie-algebras can also be constructed for each
of the supersymmetric families of matrix lie-algebras. From the super-algebra
$Open(M(L/K, {\Bbb C}))$ a Lie-product is defined using the anticommutator,
\begin{equation}
               \Phi_1 \wedge \Phi_2 = \Phi_1 \Phi_2 - \Phi_2  \Phi_1
\end{equation}
Then the lie product for elements of the representation will be anticommuting.
\begin{equation}
                     \Phi_1 \wedge \Phi_2 = - \Phi_2 \wedge \Phi_1
\end{equation}
But because of the commutation/anti-commutation relations on the components
the Lie product of two odd base elements must be symmetric instead of
anti-symmetric. I.e.
\begin{equation}
                     A \wedge B = AB - (-1)^{par(A)par(B)} BA
\end{equation}
This defines $open(gl(L/K, {\Bbb C})$.

A representation of a reduced Lie sub-algebra $open(u(L/K))$ is defined as
those elements which satisfy,
\begin{equation}
                     \Phi^\dagger = -\Phi
\end{equation}

The scalar product is now defined by
\begin{equation}
                     \Phi_1 \bullet \Phi_2 = Tr(\Phi_1^\dagger \Phi_2)
\end{equation}

This product is an invariant for the group $open(u(L/K))$ but is not
positive definite because of the extra minus sign in the trace. Only
a quantum model can be defined.

Actions for a model based on this representation are also the same as
before. In general the action may contain any powers in the algebra squared
with the scalar product.
\begin{equation}
   S  =  g_1 \Phi \bullet \Phi + g_2 \Phi^2 \bullet \Phi^2 +
                                 g_3 \Phi^3 \bullet \Phi^3 + \ldots
\end{equation}
This supersymmetric generalisation is an analogue of the supersymmetric
generalisation of matrix models already described.

It is possible that interesting physics exists in these models in
a large $L,K$ double scaling limit with the constants $g_i$ scaled as
functions of $N$.


\section*{Discrete Closed String Groups}

In continuum string theory the closed string field theories are often
considered to be of more physical interest but are also harder to construct.
The same applies to event symmetric closed string models.

It is possible to construct Closed String algebras in which the base
elements are cyclically symmetric. The extensions use a basis of closed
discrete strings which will be written with square brackets to distinguish
them from the open strings. When they are shifted cyclically a sign is
introduced if they are even length i.e.,
\begin{equation}
[a, b] = - [b, a]\\
\end{equation}
\begin{equation}
[a, b, c] = [a, b, c]\\
\end{equation}
\begin{equation}
[a, b, c, d] = - [b, c, d, a]\\
etc.
\end{equation}
Some strings of odd length must be excluded because of this sign rule e.g.
\begin{equation}
        [1,1] = [1,2,3,1,2,3] = 0
\end{equation}
The base elements are multiplied by identifying common sequences
in opposite sense within them. E.g.
\begin{equation}
        [1,2,3,4][5,3,2,7] = - [1,2,2,7,5,4] + [1,7,5,3,3,4] - [1,7,5,4]
\end{equation}
The sign for such a multiplication is chosen so that when the matching
segments are moved to the end of the first string and the beginning of the
second it is positive. For locality, when two strings have no points in common
the product is zero and the whole of a string is not cancelled against
part of another. This algebra is non-associative e.g.
\begin{equation}
                [1,2]([2,3][3,4,1]) = [1,4,1] - [2,2,4]\\
                ([1,2][2,3])[3,4,1] = [1,4,1] - [3,3,4]
\end{equation}
Because of this non-associativity we can not be sure that defining a
Lie-product as the anticommutator will satisfy the Jacobi identity.

If a string $A$ contains a piece $X$ and a string $B$ contains the
same piece reversed, i.e. $X^T$ then we can write,
\begin{equation}
                 A = a \circ X \\
		 B = X^T \circ b
\end{equation}
The circle symbol is used to mean joining pieces of strings. The term
in the multiplication which involves the cancellation of X can be written,
\begin{equation}
             (AB)_X = ((a \circ X)(X^T \circ b))_X\\
	            = a \circ b
\end{equation}
The full product can be written
\begin{equation}
                AB = \sum_X (AB)_X
\end{equation}
It can be checked that,
\begin{equation}
               (AB)_X = (-1)^{len(A) len(B) + len(X)} (BA)_{X^T}
\end{equation}
Where $len(A)$ is the number of events in a string or piece of string.
We can try to find a Lie-product which might take the form
\begin{equation}
                A \wedge B = \sum_X s(A,B,X) (AB)_X
\end{equation}
Where $s(A,B,X)$ are some form factors which must be determined to
fulfill the graded commutation and Jacobi identities which are,
\begin{equation}
                A \wedge B = (-1)^{par(A) par(B)} B \wedge A\\
     (-1)^{par(A) par(C)} (A \wedge B) \wedge C + [ cycle A,B,C ] = 0
\end{equation}
(The definition of parity of a string $par(A)$ has not yet been given.)
One way to ensure the correct commutation relations is to take
\begin{equation}
       s(A,B,X) = (1 - (-1)^{len(A) len(B) + len(X) + par(A) par(B)}) t(A,B)\\
                    t(A,B) = t(B,A)
\end{equation}
The Jacobi Identity is more difficult. In general the three strings will
have various pieces in common but could be decomposed as,
\begin{equation}
                    A = a \circ Y \circ b \circ X\\
		    B = c \circ X^T \circ d \circ Z\\
		    C = e \circ Z^T \circ f \circ Y^T
\end{equation}
With this decomposition the double product breaks into two terms and we
can write,
\begin{equation}
   (A \wedge B) \wedge C = \sum_{YX} s(A,B,X)s(AB,C,Y)((AB)_X C)_Y + \\
                           \sum_{XZ} s(A,B,X)s(AB,C,Z)((AB)_X C)_Z
\end{equation}
The following identities can be established
\begin{equation}
                   ((AB)_X C)_Z = (A (BC)_Z)_X\\
                   ((AB)_X C)_Y = (-1)^{len(B) len(C)} ((AC)_Y B)_X
\end{equation}
With this substituted into the Jacobi identity it is evident that a clear
solution is given by identifying the parity of a string with the parity
of its length and defining,
\begin{equation}
       s(A,B,X) = 1 - (-1)^{len(X)}
\end{equation}
In other words only terms with cancellation of odd length pieces are included
in the Lie-Product. (The only obvious way to get a non-abelian non-super
algebra is to restrict to the algebra generated by even length strings.)

This defines a Lie-product satisfying the super-algebra Jacobi Identity in
a non-trivial way. The Lie-product can be written as the (anti-)commutator
of the algebraic product.
\begin{equation}
                  A \wedge B = AB - (-1)^{par(A) par(B)} BA
\end{equation}
The product is not associative and the lie-product does not act as
a differential operator on the algebra satisfying the graded Leibnitz rule.
i.e.
\begin{equation}
      (AB) \wedge C \neq (-1)^{par(B) par(C)} (A \wedge C) B + A (B \wedge C)
\end{equation}
A counter example is
\begin{equation}
                      A = [1,2]\\
		      B = [3,4,5]\\
		      C = [5,4,1]
\end{equation}
This is unfortunate since it means that it is a little more difficult to
construct invariants using the product in the same way as was done for
open strings.

The real and complex Lie-algebras are given the names $closed(0/N,{\Bbb R})$
and $closed(0/N,{\Bbb C})$ respectively. The sub-algebras generated by the
length two strings are $so(0/N,{\Bbb R})$ and $so(N,{\Bbb C})$.

An encouraging feature of these closed groups is that the event symmetry
is included in the algebra in a sense that was not true for the open
string algebras. This is because the matrix sub-algebra acts on all parts
of the string in the appropriate fashion for the representation to be
considered as a family of tensor representations of the matrix algebra.
\begin{equation}
       \Phi = \sum \phi_{ab}(a,b) + \sum \phi_{abc}(a,b,c) + \ldots\\
             \phi_{ab} = - \phi_{ba}\\
	     \phi_{abc} = \phi_{bca} = \phi_{cab}\\
	     etc.
\end{equation}
The components of odd length strings are, of course, anticommuting Grassman
variables. A small change generated by the matrix sub-algebra gives e.g.
\begin{equation}
  \delta \phi_{abc} = \sum_d (\phi_{dbc} \phi_{da} + \phi_{adc} \phi_{db} +
                           \phi_{abd} \phi_{dc} )
\end{equation}
This is the correct transformation law for $\phi_{dbc}$ as a third rank
tensor under the group $SO(N)$ generated by $\phi_{dc}$. The corresponding
transformation for the open string lacks the middle term. The higher rank
components also transform correctly for the closed strings.
The Alternating group $A(N)$ is a sub-group of $SO(N)$ and acts to permute
events. Because of this the closed string algebra can be said to unify
space-time symmetries and gauge symmetries in a unique and powerful way.

In the event-symmetric open string models this unification appears to be
absent. This can be corrected by defining string groups which include
the closed strings and open strings together. This observation is perhaps
related to the fcat that continuum open string theories must necessarily
include closed strings.

The adjoint operator can be defined in the usual way for supersymmetric
adjoints on the complex algebra.
\begin{equation}
                     \Phi^{\dagger} = \sum i^{par(C)} \Phi^{C*} C^T
\end{equation}
The transpose of a string is its reversal. There is no ambiguity about which
event it is transposed because of the cyclic relations.
The sub-algebra of elements which are equal
to minus their adjoints can be taken and will be denoted by simply
$closed(0/N)$. This is an algebra of non-orientated closed discrete
super-strings.

To complete the construction of an event-symmetric closed string field theory
some invariants must be found. Because the components of the fundamental
representation transform as a family of tensors under the orthogonal matrix
subgroup it is necessary that any invariant must be written as contractions
over indices of tensor products. This condition is not sufficient however.

First of all we should look for a quadratic invariant and can try,
\begin{equation}
                I(\Phi) = \sum q(len(C)) \Phi^{C^T} \Phi^C
\end{equation}
with the a form factor $q(r)$ depending only on the length of the strings
(i.e. the rank of the tensor) to be determined. However, the odd terms are
identically zero due to anti-commutivity and the even part with $q(l) = 1$
is only invariant for the bosonic sub-group generated by even length strings.

The problem of finding invariants can be solved by using the adjoint matrix
representation. For each element $\Phi$ an infinite matrix $M(\Phi)$ acting
on the graded vector space of the algebra is defined with components,
\begin{equation}
                  M(\Phi)^A_B = \sum \Phi^C f^A_{BC}
\end{equation}
These matrices form a representation of the super Lie-algebra with the
graded anti-commutator as the Lie-product. Invariants can therefore be
constructed using trace and product.
\begin{equation}
                  Tr(M) = \sum M^C_C \\
                  I_n(\Phi) = Tr(M^n)
\end{equation}
The first invariant $I_1(\Phi)$ can be defined to be the extended trace,
\begin{equation}
                  CTr(\Phi) = I_1(\Phi) = Tr(M(\Phi))
\end{equation}
This receives contributions from even length palindromic strings e.g.,
\begin{equation}
                  CTr[1,2,2,1] = 1 \\
		  CTr[1,2,3,3,2,1] = -1
\end{equation}

With these invariants it is possible to define event symmetric quantum closed
string field theories.


\section*{Signature Groups}

Another class of groups closely related to the string groups is
based on sets of discrete events where the order does not matter
accept for a sign factor which changes according to the signature
of permutations,
\begin{equation}
        [a | b | c] = -[b | a | c]\\
	etc.
\end{equation}
Multiply by cancelling out any common events with appropriate sign factors.
To get the sign right, permute the events until the common ones are at the
end of the first set and at the start of the second in the opposite
sense. The elements can now be multiplied with the same rule as for the
open string. The same parity rules as for closed string apply. I.e. only
cancellations of an odd number of events is permitted.

The representations of these groups are families of fully antisymmetric
tensors. The Lie algebras are finite dimensional but it is not
immediately obvious how to define models with well behaved action
invariants other than the fermionic set with a similar action to the
fermionic closed string above.


\section*{Multi-loop String Groups}

The closed string group and the signature group are both sub-groups
of the larger multi-loop group. The base elements of this group represent
sets of closed loops. The closed loop group is the subgroup of single
loops in the multi-loop group and the signature groups correspond to sets
of loops each containing exactly one point.

The notation is chosen to be consistent with the loop and signature groups.
For example a double loop base element with one loop of length three and one
of two would be,
\begin{equation}
            [a,b,c|d,e] = - [a,b,c|e,d] = [b,c,a|d,e] = [d,e|a,b,c] etc.
\end{equation}
The sign factor is always the signature of the permutation on the events
in the string.

Antisymmetric multiplication is the obvious generalisation of multiplication
on the closed and signature groups.


\section*{Conclusions}

We started from a simple principle that physics could be described by
an event symmetric model and considered open and closed string field theory on
event-symmetric space-time as a possibility. The models which result
unifies space-time symmetries and string gauge groups in a simple elegant
way. Furthermore they can be recognised as natural extensions of random
matrix models which are known to be of interest in the non-perturbative
study of string theories.

It is possible that techniques used to study matrix models may also be
applicable to event-symmetric string theories and that their study may
provide further insight towards understanding the nature of superstring
theories and of space-time.

According to the classification of Isham \cite{Is93} the Event
Symmetric approach to quantum gravity would be a type IV scheme. A new
perspective is proposed from which, it is hoped, continuous
space-time, particle physics and quantum gravity arise. Any such scheme
is necessarily ambitious yet the concept of event symmetric space-time is
both simple and in keeping with previous attempts to quantise gravity and
unify particle physics.

The author welcomes all comments and corrections
by e-mail to phil@galilee.eurocontrol.fr



\end{document}